\documentclass{vgtc}                          

\newif\ifnotes
\notestrue 

\ifpdf
  \pdfoutput=1\relax                   
  \usepackage{graphicx}                
  \DeclareGraphicsExtensions{.pdf,.png,.jpg,.jpeg} 
\else
  \usepackage{graphicx}                
  \DeclareGraphicsExtensions{.eps}     
\fi%

\graphicspath{{figures/}{pictures/}{images/}{./}} 

\usepackage{microtype}                 
\PassOptionsToPackage{warn}{textcomp}  
\usepackage{textcomp}                  
\usepackage{mathptmx}                  
\usepackage{times}                     
\usepackage{cite}                      
\usepackage{tabu}                      
\usepackage{booktabs}                  

\onlineid{0}

\vgtccategory{Research}

\vgtcinsertpkg

\title{The 2nd MERCADO Workshop at IEEE VIS 2025: \\ Multimodal Experiences for Remote Communication Around Data Online}

\author{Wolfgang Büschel\thanks{e-mail: bueschel@acm.org}\\ %
    \scriptsize VISUS, University of Stuttgart, Germany %
\and Gabriela Molina Le\'{o}n\thanks{e-mail: leon@cs.au.dk}\\ %
     \scriptsize Aarhus University, Denmark %
\and Arnaud Prouzeau\thanks{e-mail: arnaud.prouzeau@inria.fr}\\ %
     \scriptsize Université Paris-Saclay, CNRS, Inria, France
\and Mahmood Jasim\thanks{e-mail: mjasim@lsu.edu}\\ %
     \scriptsize Louisiana State University, USA
\and Christophe Hurter\thanks{e-mail: christophe.hurter@enac.fr}\\ %
    \scriptsize  University of Toulouse, France 
\and Maxime Cordeil\thanks{e-mail: m.cordeil@uq.edu.au}\\ %
    \scriptsize University of Queensland, Australia
\and Matthew Brehmer\thanks{e-mail: mbrehmer@uwaterloo.ca}\\ %
    \scriptsize University of Waterloo, Canada
}

\abstract{
We propose a half-day workshop at IEEE VIS 2025 on addressing the emerging challenges in data-rich multimodal remote collaboration.
We focus on synchronous, remote, and hybrid settings where people take part in tasks such as data analysis, decision-making, and presentation.
With this workshop, we continue successful prior work from the first MERCADO workshop at VIS 2023 and a 2024 Shonan Seminar that followed.
Based on the findings of the earlier events, we invite research and ideas related to four themes of challenges: Tools \& Technologies, Individual Differences \& Interpersonal Dynamics, AI-assisted Collaboration, and Evaluation.
With this workshop, we aim to broaden the community, foster new collaborations, and develop a research agenda to address these challenges in future research.
Our planned workshop format is comprised of a keynote, short presentations, a breakout group session, and discussions organized around the identified challenges.

} 

\usepackage{hyperref}
\usepackage{enumitem}
\setlist{topsep=0pt, leftmargin=*}

\newcommand{\bdef}[1]{\vspace{1mm} \noindent{\textbf{#1}}}

\begin{document}


\firstsection{Workshop Proposal: Introduction}
\label{sec:intro}

\maketitle
The last years saw the normalization of remote and hybrid work, not least due to the COVID-19 pandemic.
With it came an increasing demand for effective remote communication within companies, government agencies, and educational and research institutions.
Analyzing data, making decisions grounded in this data, and presenting the outcomes to wider audiences are complex use cases that bring different subject experts, decision-makers, and other stakeholders together.

Today's popular teleconference tools such as Zoom\footnote{\url{https://zoom.us/}}, Cisco Webex Meetings\footnote{\url{https://www.webex.com/meetings.html}}, Slack Huddles\footnote{\url{https://slack.com/features/huddles}}, and Google Meet\footnote{\url{https://meet.google.com/}} have matured fast in the last few years and typically provide full solutions for multimodal communication, including video and audio conferencing, screen sharing, breakout rooms, polls, reactions, and side-channel text chat functionality.
In many cases, such tools are used together with collaborative productivity tools (e.g., Mattermost\footnote{\url{https://mattermost.com/}}, Microsoft Teams\footnote{\url{https://www.microsoft.com/microsoft-teams}}), online writing tools, mind-mapping (e.g., Miro\footnote{\url{https://miro.com/}}), and sketching tools.
Together, they support synchronous collaboration within larger, asynchronous workflows.

Despite recent advances in teleconference systems, significant gaps remain in supporting collaborative data exploration and interactive presentation in real time \cite{isenberg2011collaborative}. Most platforms still rely heavily on static screen-sharing or slide-based content, limiting participants’ ability to jointly engage with and analyze data. As Heer and Shneiderman argue, static views address only a narrow range of questions and thus must be augmented by iterative exploration and dynamic interactions for meaningful analysis \cite{Heer2012InteractiveDynamics}. More recently, Brehmer and Kosara highlight that although current tools might suffice for large-audience presentations, smaller team meetings call for more flexible, interactive data communication \cite{brehmer2021jam}. Researchers in collaborative systems have identified the need for finer-grained interaction controls, richer awareness cues, and synchronous co-creation features \cite{gutwin2002descriptive}, yet few existing tools fully address these requirements. This shortfall leaves an opportunity for new approaches that better support exploratory, data-focused collaboration.

The goals of the first MERCADO workshop at VIS 2023~\cite{brehmer2023mercadoworkshopieeevis} included assembling a corpus of inspiring examples and work to outline a design space for data-rich remote collaboration. Building on the fruitful discussions from the workshop, a follow-up NII Shonan Meeting~\cite{brehmer2024shonanseminar} in 2024 led to a structured thematic analysis of the grand challenges faced by researchers and practitioners when developing data-rich, collaborative systems for remote or hybrid settings. It is with the momentum of this meeting that we continue our investigations of this design space with this workshop proposal.

\begin{figure}[h!]
    \vspace{-2.5mm}
    \includegraphics[width=\linewidth]{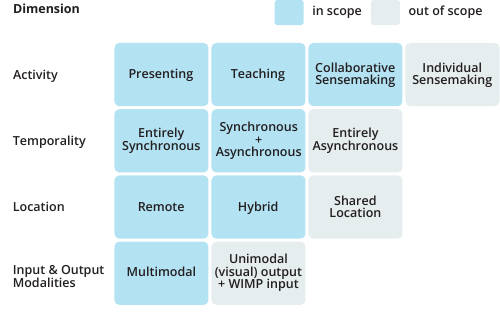}
    \vspace{-5mm}
    \caption{This diagram summarizes the continuing scope of our proposed workshop on communication and collaboration around data. Based on \cite{brehmer2023mercadoworkshopieeevis}.}
    \label{fig:scope}
    \vspace{-2.5mm}
\end{figure}

In this half-day workshop, we retain the broad scope established in the previous iteration (see \autoref{fig:scope}). The scope lies at the intersection of data visualization, human-computer interaction, computer-supported cooperative work, and online education. However, we have updated and expanded it based on insights and feedback from the prior workshop and the Shonan seminar. We continue to see this workshop as an opportunity to grow the community and address the increasing interest in new interactive experiences for synchronous and multimodal communication and collaboration around data with remote or hybrid audiences.

\section{Related Prior Workshops and Seminars}
\label{sec:prior}

Our proposed workshop is a continuation of the first MERCADO workshop at VIS 2023 \cite{brehmer2023mercadoworkshopieeevis} in Australia and the following 2024 Shonan Seminar \cite{brehmer2024shonanseminar} on \emph{Augmented Multimodal Interaction for Synchronous Presentation, Collaboration, and Education with Remote Audiences} in Japan. 
The first workshop was intended to start building a community around multimodal and data-rich collaborative experiences.
The work presented led to multiple success stories: at least two publications followed from accepted workshop submissions \cite{leon25wall, cordeil25}, and another paper is currently under review.

The first workshop and the Shonan seminar brought together an international group of experts from different areas, including information visualization, immersive analytics, multimodal interaction, accessibility, human-centered AI, and computer-supported cooperative work. This pool of experts, which goes beyond the VIS community and academia, will serve as a starting point for the program committee of our workshop.
Our organizing team includes organizers of the first workshop and seminar participants.

A main goal of these events was to identify emerging challenges in data-rich collaborations. These challenges, shortly described in the seminar report \cite{brehmer2024shonanseminar}, take into account the recent technological developments since the planning of the first workshop, such as the pervasive use of large language models (e.g., \cite{Anil2023gemini, guo2024deepseekcoder, gunter2024appleintelligence}).
While the prior events focused on identifying the challenges, our proposed workshop will focus on addressing them. 
We will invite submissions on approaches to investigate and develop solutions and will combine presentations with interactive activities to support further development.
Moreover, we will broaden the community by involving researchers, practitioners, and educators based in Europe, providing a dedicated space for collaborations. 

The first workshop and seminar have significantly contributed to advancing our understanding and outlining a roadmap for action in this domain. While some progress has been made, substantial work remains. In this second instance of the MERCADO workshop series, we aim to broaden the scope and deepen the exploration of key challenges.

Prior workshops at IEEE VIS have partially covered some of the topics we plan to address, such as the use of generative AI at the data storytelling workshop \cite{lan2024gen4dsworkshop} and evaluation at BELIV \cite{beliv24}, but ours is the only one focusing on data-rich multimodal collaboration.

\begin{figure}[h!]
    \vspace{-2.5mm}
    \includegraphics[width=\linewidth]{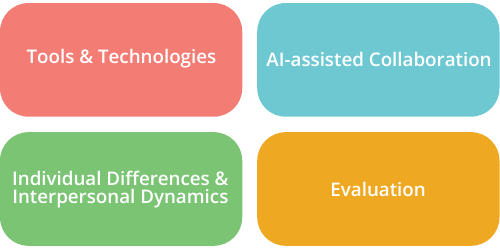}
    \vspace{-5mm}
    \caption{The four main themes of challenges along which we structure our workshop.}
    \label{fig:events}
    \vspace{-2.5mm}
\end{figure}

\section{Workshop Goals and Research Questions}
\label{sec:goals}

Our proposed workshop will focus on research and technologies aiming to address the main challenges of future remote communication and collaboration around data.
With this workshop, we would also like to grow the community around this area of research, especially in Europe, help spawn and advance research collaborations by connecting like-minded researchers, and foster deeper international networking.
With the results of this workshop, we aim to contribute to several academic communities, including those affiliated with the IEEE VGTC (VIS, ISMAR, Eurovis, PacificVis) and ACM SIGCHI (CHI, CSCW, UIST, ISS).
Based on the outcomes of the prior workshop and seminar, we will invite participants to submit ideas and positions relating to one or more of the following four categories of future research challenges:

\begin{itemize}
    \item \emph{Techniques and Technologies}: The development of future systems for data-rich remote collaboration is dependent on fundamental technological questions. Which technologies will elevate hybrid and remote settings to reach or even surpass face-to-face meetings? What are novel visualization and interaction techniques most suitable for such scenarios?
    \item \emph{Individual Differences and Group Dynamics}: One of the main advantages of remote and hybrid communication is the chance to bring together various experts from multiple fields with different personal preferences. How can we support people from diverse backgrounds and with their own goals, needs, and roles? How can we scale data-rich collaborative use cases to many participants?
    \item \emph{AI-Assisted Collaboration}: The influence that AI technologies have on our society is undeniable. One of the core challenges of future research in the area of remote communication \& collaboration with data is to integrate such technologies while mitigating potential risk factors regarding reliability, explainability, and privacy.
    How will AI assistants and agents shape future collaborative data analysis? How can we ensure trustworthiness and limit bias?
    \item \emph{Evaluation}: More research is needed to understand the power as well as the limitations of remote and hybrid data-rich communication \& collaboration. Yet, the complexity of current and future systems and a lack of established baselines and methods means that investigating these systems is difficult. What are the benefits of hybrid collaboration for data analysis in different use cases? Which tools or methods are best suited to evaluate this?
\end{itemize}

Overall, our aim is to identify and curate timely research objectives to support the research community in addressing the major challenges of data-rich communication \& collaboration in remote and hybrid settings. As such, while we focus on the topics above, we are also open to additional exciting ideas addressing challenges, bridging or going beyond these four themes.
In particular, we invite participants to submit work that extends and investigates areas identified as needing further research in the field of Multimodal Experiences for Remote Communication Around Data Online: accessibility, inclusivity, comparisons between synchronous and asynchronous communication, brain-computer interfaces, human-machine collaboration, affective computing, embodied cognition, and more.

\section{Format and Planned Activities}
\label{sec:format}

This workshop is planned as a half-day event.
Given our topic, we intend to offer a fully hybrid workshop that is open to remote attendance.
Assuming a schedule similar to recent VIS conferences, we anticipate two 75-minute workshop sessions on either side of a 30-minute coffee break.
Overall, the tentative schedule is as follows:

\begin{itemize}
    \item 
    \textbf{Session 1}: Introduction to the workshop goals and activities (15 min), followed by lightning talks and brief discussions by participants (30 min), and a keynote by an invited speaker (30 min).
    \item
    \textbf{Break}: Coffee break and, in parallel, short live demos by the participants and the workshop organizers (30 min).
    \item
    \textbf{Session 2}: Breakout groups to ideate on how to address the main challenges along the four proposed themes (45 min), reports by the groups, and a short discussion (20 min).
    \item
    \textbf{Conclusion and Call to Action}:
    To conclude the workshop, we will summarize the progress made so far and emphasize that the topic is both broad and increasingly important, deserving continued attention and effort. We will invite the audience to reflect on potential individual and collective actions that could help expand and disseminate the knowledge developed in this area. As examples---though not limited to these---we will highlight applications in education, healthcare, and the development of effective, data-driven collaborative decision-making tools (10 min).
\end{itemize}

As VIS 2025 will be held in Austria, we see this workshop as an opportunity to invite a local practitioner to inspire attendees. However, we are also open to a remote talk.
With one of our workshop goals being to strengthen the community, we plan to provide a longer-lasting platform for discussion and collaboration using a service such as Discord or Mattermost, and we will organize an informal event for networking later in the conference. We will discuss the details with the workshop participants during the second session.

\section{Accommodations and Contingencies}
\label{sec:accommodations}

As described above, following the theme of remote and hybrid collaboration in this workshop, we would like to offer hybrid participation regardless of the overall VIS 2025 conference format.
If this workshop proposal is accepted, we will gladly initiate conversations with the VIS organizing committee at an early stage to ensure a successful hybrid event.
To run the workshop, we require a standard VIS workshop room with AV equipment (microphones for the presenter and for questions in the audience, projector).
Considering the planned hybrid participation, we require a stable WiFi connection with suitable bandwidth for video conferencing.
Participants will join with their laptops, and presenters will use the workshop org laptop to present on the projector and in Zoom. As an online collaboration tool, we plan on using a Miro\footnote{\url{https://miro.com}} page created with a pro account, allowing both on-site and remote attendees to participate. Talks may be recorded with Zoom if needed.

\section{Publication Plan}
\label{sec:publication}

We invite prospective participants to submit a short paper (up to four pages) in the style of work-in-progress reports, case studies, or short provocation or reflection statements.
We encourage submitting accompanying videos.
To accommodate different forms of contributions, we will accept submissions using the IEEE VGTC Conference Style Template format, in a pictorial format similar to submissions made to the VIS Arts Program\footnote{\url{https://visap.net}}, or in an interactive article format similar to submissions to the VISxAI workshop\footnote{\url{https://visxai.io}}.
We will host accepted submissions on the workshop's website (\texttt{\href{https://sites.google.com/view/mercadoworkshop/vis2025}{sites.google.com/view/mercadoworkshop/vis2025}}). 
Furthermore, we will recommend authors to share their publications using open-access services. The accepted submissions will not be considered archival, and we will encourage the reuse of content in follow-up publications. Submissions will not be anonymous; authors must include their full names, emails, and affiliations. Lastly, at least one author per accepted submission must register for the conference.

\section{Call for Participation and Timeline}
\label{sec:participation}


If accepted, we plan to announce our call for participation in time for CHI (April 26) and PacificVis (April 22), advertising on IEEE and ACM mailing lists as well as on social media.
We target a submission deadline following the first-round notification of full VIS papers (early June), with a short review period concluding with author notifications in mid-July.   
Our program committee will be recruited from the 29 attendees of the 2024 NII Shonan meeting, which includes experts in several adjacent fields of research (e.g., CSCW, XR, accessibility).
In addition to overall quality, our main selection criterion will be the relevance of submissions to our topic of communicating and collaborating around data with remote and hybrid audiences across the four proposed challenge themes.

\section{Organizers}
\label{sec:organizers}

\bdef{Wolfgang Büschel} is a postdoctoral researcher at VISUS, University of Stuttgart, Germany, where he specializes in immersive and situated analytics. Previously, he was with the Interactive Media Lab of TUD Dresden University of Technology. There, he also did his PhD on spatial interaction with immersive visualizations. For his thesis, he received the Dissertation Award of Commerzbank's Dresden Cultural Foundation. His research interests include the situated analysis of interaction data~\cite{bueschel2021miria}, immersive data storytelling~\cite{mendez2025immersive}, and the use of mobile devices to support 3D interaction~\cite{krug2022clear,bueschel2019investigating}. Wolfgang Büschel has published more than 30 peer-reviewed papers and received several best paper awards. His works have been cited over 1200 times.
\textbf{Website}: \texttt{\href{https://wolfgang-bueschel.de/}{wolfgang-bueschel.de}}.

\bdef{Gabriela Molina Le\'{o}n} is a postdoctoral research fellow at Aarhus University, Denmark, with a focus on collaborative visual analytics.
Previously, she did her PhD at the Institute of Information Management Bremen at the University of Bremen. As part of her research, she has co-authored several publications on the design and development of multimodal systems to support data experts in performing visual data exploration \cite{molina2022mobile} and collaborative sensemaking \cite{leon25wall}. Her research lies at the intersection of visualization, computer-supported cooperative work, and human-computer interaction. She has co-organized workshops at CSCW 2021 \cite{cjcollective21cscw} and CLIHC 2021 \cite{cjcollective21clihc}, and is also one of the organizers of the Data Visualization Bremen meetup.
\textbf{Website}: \texttt{\href{https://gmleon.github.io/}{gmleon.github.io}}.

\bdef{Arnaud Prouzeau} is a human-computer interaction and visualization researcher at Inria in Paris and previously Bordeaux, France. Prior to Inria, he was a research fellow at Monash University, Australia. He brings an array of expertise from various projects involving the exploration of visualization with tangible interfaces \cite{marlier2025exploring}, virtual and augmented reality \cite{liu2023datadancing}, and collaborative systems \cite{lee2020shared}. He led Monash University's Cognitive Building Initiative in collaboration with Honeywell, aiming to integrate immersive analytics and machine learning to support building management. Currently, he is leading the 'ICARE' project\footnote{\url{https://icare.inria.fr/}}, a multidisciplinary initiative focusing on the design of collaborative environments for hands-on educational activities.
\textbf{Website}: \texttt{\href{https://www.aprouzeau.com/}{aprouzeau.com}}.

\bdef{Mahmood Jasim} is an assistant professor at Louisiana State University (LSU), USA, where he specializes in building decision support systems to inform collaborative and personalized decision-making. Before joining LSU, he completed his PhD at the University of Massachusetts Amherst. His dissertation focused on inclusive collection and analysis of public-generated data for informed decision-making. His research intersects across information visualization, human-computer interaction, and social computing to build and deploy systems to inform collaborative analysis and sensemaking\cite{jasim2023communityclick} and personalized decisions~\cite{jasim2021communitypulse, jasim2022supporting}. His published research received awards from CSCW 2020, DIS 2021, and EuroVIS 2022. 
\textbf{Website}: \texttt{\href{https://csc.lsu.edu/~mjasim}{csc.lsu.edu/\textasciitilde mjasim}}.

\bdef{Christophe Hurter} is a Professor of Computer Science at ENAC (the French Civil Aviation University) and co-founder of Avrio MedTech, a company focused on medical and epileptic data analysis through transparent AI. His research spans Human-Computer Interaction, Human-Centered AI, and Interactive Data Visualization, with applications in aviation, air traffic control, and healthcare. His work bridges theoretical advances and practical solutions, emphasizing explainability, user trust, and immersive data exploration in critical domains. He was an organizer of Shonan Seminar \#213~\cite{brehmer2024shonanseminar} and the first MERCADO workshop at VIS 2023~\cite{brehmer2023mercadoworkshopieeevis}.
\textbf{Website}: \texttt{\href{http://recherche.enac.fr/~hurter/}{recherche.enac.fr/\textasciitilde hurter}}.

\bdef{Maxime Cordeil} is a Senior Lecturer at the University of Queensland, Australia. Dr. Cordeil has been recognized as Australia’s top researcher in computer graphics (2021, 2022). His research focuses on human-computer interaction, data visualisation, and analytics. He has published over 60 journal and conference in top venues such as ACM CHI, IEEE VIS or IEEE VR. Dr. Cordeil is a key international member of the Immersive Analytics community of researchers and has organised several workshops on the topic of Immersive Analytics (``IA Workshop series'' at VIS 2017, CHI 2018, CHI 2019, CHI 2020, and CHI 2022). The activities of the IA community focus on designing and evaluating the future graphical user interfaces for data analysis in Virtual / Augmented Reality. He also co-organized the first MERCADO workshop at VIS 2023~\cite{brehmer2023mercadoworkshopieeevis}.  \textbf{Website}: \texttt{\href{https://sites.google.com/view/cordeil/home}{sites.google.com/view/cordeil}}.

\bdef{Matthew Brehmer} is an assistant professor with the School of Computer Science at the University of Waterloo, where he directs the \href{https://www.ubixgroup.ca/}{Ubietous Information Experiences Group}. Prior to joining the University of Waterloo, he was a Lead Researcher at Tableau. His work has been recognized with Best Paper (IEEE VIS), Honorable Mention (IEEE VIS, ACM UIST), and Test-of-Time Awards (IEEE VIS). He was an organizer of Shonan Seminar \#213~\cite{brehmer2024shonanseminar}, the first MERCADO workshop at VIS 2023~\cite{brehmer2023mercadoworkshopieeevis}, the \href{https://mobilevis.github.io/chi18workshop.html}{MobileVis workshop at CHI 2018}, and the \href{https://ieeevis.org/year/2021/info/visinpractice}{VisInPractice event at IEEE VIS} between 2018 and 2021. In 2022, he was elected to the VIS Executive Committee (VEC) and appointed to the IEEE Visualization and Computer Graphics Technical Community (VGTC) Executive Committee. \textbf{Website}: \texttt{\href{https://mattbrehmer.ca/}{mattbrehmer.ca}}.

\bibliographystyle{abbrv-doi}

\bibliography{main}

\begin{thebibliography}{10}

\bibitem{beliv24}
2024 {IEEE} evaluation and beyond - methodological approaches for visualization.
\newblock In {\em Proc.\ BELIV}. IEEE, 2024. doi: {{%
10\discretionary{/}{%
}{/}g9bxsc}}


\bibitem{Anil2023gemini}
R.~Anil, S.~Borgeaud, Y.~Wu, J.~Alayrac, J.~Yu, R.~Soricut, J.~Schalkwyk, A.~M. Dai, A.~Hauth, K.~Millican, D.~Silver, S.~Petrov, M.~Johnson, and et~al.
\newblock Gemini: {A} family of highly capable multimodal models.
\newblock {\em CoRR}, 2023. doi: {{%
10\discretionary{/}{%
}{/}g9bxr8}}


\bibitem{brehmer2023mercadoworkshopieeevis}
M.~Brehmer, M.~Cordeil, C.~Hurter, and T.~Itoh.
\newblock The {MERCADO} workshop at {IEEE VIS} 2023: Multimodal experiences for remote communication around data online, 2023.

\bibitem{brehmer2024shonanseminar}
M.~Brehmer, M.~Cordeil, C.~Hurter, and T.~Itoh.
\newblock Augmented multimodal interaction for synchronous presentation, collaboration, and education with remote audiences, 2024.
\newblock NII Shonan Report \#213, \href{https://shonan.nii.ac.jp/seminars/213/}{https://shonan.nii.ac.jp/seminars/213/}.

\bibitem{brehmer2021jam}
M.~Brehmer and R.~Kosara.
\newblock From jam session to recital: Synchronous communication and collaboration around data in organizations.
\newblock {\em IEEE TVCG}, 28(1), 2022. doi: {{%
10\discretionary{/}{%
}{/}gsvfm5}}


\bibitem{bueschel2021miria}
W.~B\"{u}schel, A.~Lehmann, and R.~Dachselt.
\newblock {MIRIA}: A mixed reality toolkit for the in-situ visualization and analysis of spatio-temporal interaction data.
\newblock In {\em Proc.\ CHI}. ACM, 2021. doi: {{%
10\discretionary{/}{%
}{/}gksmrg}}


\bibitem{bueschel2019investigating}
W.~B\"{u}schel, A.~Mitschick, T.~Meyer, and R.~Dachselt.
\newblock Investigating smartphone-based pan and zoom in {3D} data spaces in augmented reality.
\newblock In {\em Proc.\ MobileHCI}. ACM, 2019. doi: {{%
10\discretionary{/}{%
}{/}gtdktp}}


\bibitem{cjcollective21clihc}
C.~J. Collective, D.~de~Castro~Leal, G.~Molina~Le\'{o}n, J.~F. Maestre, K.~Williams, M.~Wong-Villacres, P.~Reynolds-Cu\'{e}llar, S.~K. Oswal, T.~Cerratto~Pargman, and V.~Sharma.
\newblock Citational practices: Interrogating hegemonic knowledge structures in computing research in latin america.
\newblock In {\em Proc.\ CLIHC}. ACM, 2021. doi: {{%
10\discretionary{/}{%
}{/}gn2zqh}}


\bibitem{cjcollective21cscw}
C.~J. Collective, G.~Molina~Le\'{o}n, L.~Kirabo, M.~Wong-Villacres, N.~Karusala, N.~Kumar, N.~Bidwell, P.~Reynolds-Cu\'{e}llar, P.~P. Borah, R.~Garg, S.~K. Oswal, T.~Chuanromanee, and V.~Sharma.
\newblock Following the trail of citational justice: Critically examining knowledge production in hci.
\newblock In {\em Proc.\ CSCW}, pp. 360--363. ACM, 2021. doi: {{%
10\discretionary{/}{%
}{/}g9bxr6}}


\bibitem{cordeil25}
M.~Cordeil, A.~Servais, G.~Truong, T.~Dwyer, D.~Vyas, and C.~Hurter.
\newblock The presenter in the browser: Design and evaluation of human interactive overlays with web content.
\newblock {\em Multimodal Technol. Interact.}, 9(2), 2025. doi: {{%
10\discretionary{/}{%
}{/}g9bxsg}}


\bibitem{gunter2024appleintelligence}
T.~Gunter, Z.~Wang, C.~Wang, R.~Pang, A.~Narayanan, A.~Zhang, B.~Zhang, C.~Chen, C.-C. Chiu, D.~Qiu, D.~Gopinath, D.~A. Yap, D.~Yin, et~al.
\newblock Apple intelligence foundation language models, 2024. doi: {{%
10\discretionary{/}{%
}{/}g9bxr9}}


\bibitem{guo2024deepseekcoder}
D.~Guo, Q.~Zhu, D.~Yang, Z.~Xie, K.~Dong, W.~Zhang, G.~Chen, X.~Bi, Y.~Wu, Y.~K. Li, F.~Luo, Y.~Xiong, and W.~Liang.
\newblock Deepseek-coder: When the large language model meets programming -- the rise of code intelligence, 2024. doi: {{%
10\discretionary{/}{%
}{/}gtg8cw}}


\bibitem{gutwin2002descriptive}
C.~Gutwin and S.~Greenberg.
\newblock A descriptive framework of workspace awareness for real-time groupware.
\newblock {\em Computer Supported Cooperative Work (CSCW)}, 11:411--446, 2002. doi: {{%
10\discretionary{/}{%
}{/}c43wsg}}


\bibitem{Heer2012InteractiveDynamics}
J.~Heer and B.~Shneiderman.
\newblock Interactive dynamics for visual analysis.
\newblock {\em Commun. ACM}, 55(4):45–54, Apr. 2012. doi: {{%
10\discretionary{/}{%
}{/}gd2wmh}}


\bibitem{isenberg2011collaborative}
P.~Isenberg, N.~Elmqvist, J.~Scholtz, D.~Cernea, K.-L. Ma, and H.~Hagen.
\newblock Collaborative visualization: Definition, challenges, and research agenda.
\newblock {\em Information Visualization}, 10(4), 2011. doi: {{%
10\discretionary{/}{%
}{/}b93p9h}}


\bibitem{jasim2022supporting}
M.~Jasim, C.~Collins, A.~Sarvghad, and N.~Mahyar.
\newblock Supporting serendipitous discovery and balanced analysis of online product reviews with interaction-driven metrics and bias-mitigating suggestions.
\newblock In {\em Proc.\ CHI}. ACM, 2022. doi: {{%
10\discretionary{/}{%
}{/}kr28}}


\bibitem{jasim2021communitypulse}
M.~Jasim, E.~Hoque, A.~Sarvghad, and N.~Mahyar.
\newblock {CommunityPulse}: Facilitating community input analysis by surfacing hidden insights, reflections, and priorities.
\newblock In {\em Proc.\ DIS}, p. 846–863. ACM, 2021. doi: {{%
10\discretionary{/}{%
}{/}g5jp7s}}


\bibitem{jasim2023communityclick}
M.~Jasim, A.~Sarvghad, and N.~Mahyar.
\newblock {CommunityClick-Virtual}: Multi-modal interactions for enhancing participation in virtual meetings, Sep 2023. doi: {{%
10\discretionary{/}{%
}{/}g9bxr7}}


\bibitem{krug2022clear}
K.~Krug, W.~Büschel, K.~Klamka, and R.~Dachselt.
\newblock {CleAR Sight}: Exploring the potential of interacting with transparent tablets in augmented reality.
\newblock In {\em Proc.\ ISMAR}, pp. 196--205. IEEE, 2022. doi: {{%
10\discretionary{/}{%
}{/}mbw6}}


\bibitem{lan2024gen4dsworkshop}
X.~Lan, L.~Yang, Z.~Wang, Y.~Wang, D.~Shi, and S.~Carpendale.
\newblock Gen4ds: Workshop on data storytelling in an era of generative {AI}, 2024. doi: {{%
10\discretionary{/}{%
}{/}g9bxsb}}


\bibitem{lee2020shared}
B.~Lee, X.~Hu, M.~Cordeil, A.~Prouzeau, B.~Jenny, and T.~Dwyer.
\newblock Shared surfaces and spaces: Collaborative data visualisation in a co-located immersive environment.
\newblock {\em IEEE TVCG}, 27(2):1171--1181, 2020. doi: {{%
10\discretionary{/}{%
}{/}ghgt5v}}


\bibitem{liu2023datadancing}
J.~Liu, B.~Ens, A.~Prouzeau, J.~Smiley, I.~K. Nixon, S.~Goodwin, and T.~Dwyer.
\newblock {DataDancing}: An exploration of the design space for visualisation view management for {3D} surfaces and spaces.
\newblock In {\em Proc.\ CHI}. ACM, 2023. doi: {{%
10\discretionary{/}{%
}{/}g9bxsf}}


\bibitem{marlier2025exploring}
M.~Marlier, N.~Renoir, M.~Hachet, and A.~Prouzeau.
\newblock Exploring interactions with tangible and actuated tokens on a shared tabletop for railway traffic management control centres.
\newblock In {\em Proc.\ TEI}. ACM, 2025. doi: {{%
10\discretionary{/}{%
}{/}g9bxsd}}


\bibitem{leon25wall}
G.~Molina~Le{\'o}n, A.~Bezerianos, O.~Gladin, and P.~Isenberg.
\newblock Talk to the wall: The role of speech interaction in collaborative visual analytics.
\newblock {\em IEEE TVCG}, 31(1):941--951, 2025. doi: {{%
10\discretionary{/}{%
}{/}g88nds}}


\bibitem{molina2022mobile}
G.~Molina~Le{\'o}n, M.~Lischka, W.~Luo, and A.~Breiter.
\newblock Mobile and multimodal? {A} comparative evaluation of interactive workplaces for visual data exploration.
\newblock {\em Computer Graphics Forum}, 41(3):417--428, 2022. doi: {{%
10\discretionary{/}{%
}{/}g88ncq}}


\bibitem{mendez2025immersive}
J.~Méndez, W.~Luo, R.~Rzayev, W.~Büschel, and R.~Dachselt.
\newblock Immersive data-driven storytelling: Scoping an emerging field through the lenses of research, journalism, and games.
\newblock {\em IEEE TVCG}, pp. 1--13, 2025. doi: {{%
10\discretionary{/}{%
}{/}n9dj}}


\end{thebibliography}
\end{document}